\documentclass[aps,twocolumn,groupedaddress,amsmath,amssymb,showpacs]{revtex4-1}

\usepackage{graphicx}
\usepackage{ae}
\usepackage{color}
\usepackage{bm}
\usepackage{float}

\newcommand{\be}{\begin{equation}} \newcommand{\ee}{\end{equation}}
\newcommand{\ord}{\mathcal{O}}
\newcommand{\lang}{\left\langle} \newcommand{\rang}{\right\rangle}

\begin{document}
\title{On the effect of intermittency of turbulence on the parabolic relation between skewness and kurtosis in magnetized plasmas}

\author{D\'avid Guszejnov}
\email[]{guszejnov@reak.bme.hu}
\affiliation{Department of Nuclear Techniques, Budapest University of Technology and Economics, Association EURATOM, M\H{u}egyetem rkp. 9., H-1111 Budapest, Hungary}

\author{Attila Bencze}
\affiliation{MTA Wigner RCP, EURATOM Association, PO Box 49, H-1525 Budapest, Hungary}

\author{S\'andor Zoletnik}
\affiliation{MTA Wigner RCP, EURATOM Association, PO Box 49, H-1525 Budapest, Hungary}

\author{N\'ora Laz\'anyi}
\affiliation{Department of Nuclear Techniques, Budapest University of Technology and Economics, Association EURATOM, M\H{u}egyetem rkp. 9., H-1111 Budapest, Hungary}

\date{\today}

\begin{abstract}

 This paper is aimed to contribute to the scientific discussions that have been triggered by the experimental observation of a quadratic relation between the kurtosis and skewness of turbulent fluctuations present in fusion plasmas and other nonlinear physical systems. In this paper we offer a general statistical model which attributes the observed $K = aS^2 + b$ relation to the varying intermittency of the experimental signals. The model is a two random variable model constructed to catch the essential intermittent feature of the real signal. One of the variables is the amplitude of the underlying intermittent event (e.g. turbulent structure) while the other is connected to the intermittency level of the system. This simple model can attribute physical meaning to the $a$ and $b$ coefficients, as they characterize the spatio-temporal statistics of intermittent events. By constructing a particle-conserving Gaussian model for the underlying coherent structures the experimentally measured $a$ and $b$ coefficients could be adequately reproduced.
 
\end{abstract}

\maketitle

\section{Introduction}

\begin{figure}[h]
\begin {center}
\includegraphics[width=0.9 \linewidth]{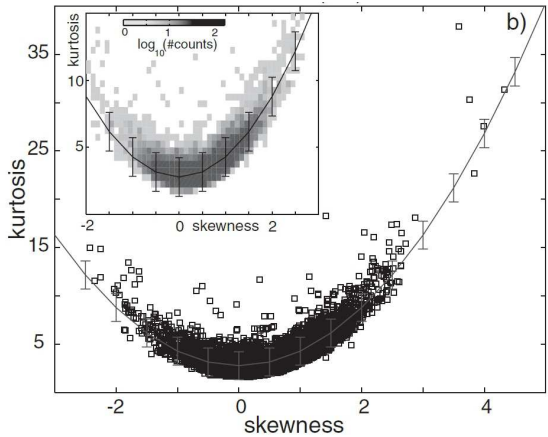}
\caption{Measured skewness and kurtosis of TORPEX signals \cite{torpex}. Figure reprinted with permission from B. Labit et al., Physical Review Letters, 98, 255002 2007 (http://link.aps.org/abstract/PRL/v98/e255002), Fig. 1. Copyright 2007 by the American Physical Society. }\label{fig:torpex_SK}
\end {center}
\end{figure}

In 2007 Labit et al. published an analysis of Langmuir probe measurements in the TORPEX device \cite{torpex}. They calculated the third and fourth normalized central moments (\textit{skewness} and \textit{kurtosis}) of the signals and found a strong parabolic relation (Fig. \ref{fig:torpex_SK}):
\be
\label{eq:exp_parabola}
K=2.78+1.5 S^2.
\ee

The authors investigated the possible physical causes of this quadratic relation and found that the correlation vanishes if the frequency range characteristic to the drift-interchange instabilities is filtered out, which implies that in the case of plasmas coherent turbulent structures are responsible for this phenomenon. They also found that the generalized Beta distribution -- and in some cases its special case, the Gamma distribution -- is a very good fit for the experimental signals. Nevertheless, no physical motivation was provided to the signals having a Beta distribution, apart from them being non-Gaussian. Similar relations were found in other devices \cite{Labit2, Sattin_more_device}, although with different coefficients. The Beta distribution also proved to be a good fit for these experimental data.

\begin{figure}[h]
\begin {center}
\includegraphics[width=0.9 \linewidth]{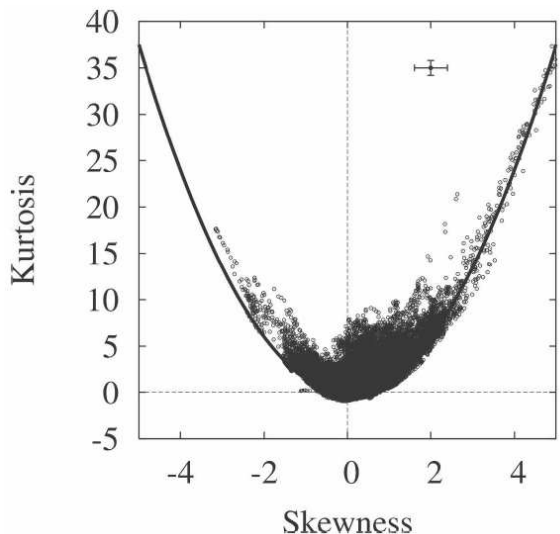}
\caption{Measured skewness and kurtosis of daily sea surface temperature (SST). Figure reprinted with permission from Ref. \cite{SS}, Fig. 3, Copyright 2007 by the American Meteorological Society.}\label{fig:SS_SK}
\end {center}
\end{figure}

Meanwhile a very similar relation was found in a very different physical system. Sura and Sardeshmukh in 2007 found that the daily \textit{sea surface temperature} (SST) exhibits a similar parabolic relation between skewness and kurtosis (Fig. \ref{fig:SS_SK}) \cite{SS}. The lack of similarity between these systems implied that the cause of the parabolic relation is a model independent, universal property.

In 2008 Krommes presented a possible solution, which argued that the governing equations of the plasma turbulence and the SST fluctuations take similar form in a Langevin representation \cite{krommes}. Krommes also showed that this common Langevin equation leads to a parabolic relation between skewness and kurtosis. However, there are some deficiencies in this model. The Langevin representation employs Gaussian fields, while it is well known that turbulence is a highly nonlinear phenomenon, so the fields should be strongly non-Gaussian. Further problems arise when one considers that the SST fluctuations are dominantly excited by external sources (e.g. wind) that are strongly coupled to the system, while plasma density fluctuations are mainly caused by the internal dynamics of the system.

In 2009 Sattin et al. proposed a universal explanation to the parabolic relation in case of both the TORPEX fluctuations and the sea surface temperature \cite{Sattin}. They showed that every physical system that fulfills several not very restrictive conditions will in fact have a parabolic relation between its skewness and kurtosis. The conditions are the following:

\begin{enumerate}
	\item The system can be described with stochastic differential equations (probably only Markovian systems, although this is not stated in the paper).
	\item There exists a dominant parameter, which influences the expected skewness and kurtosis values.
	\item Under very weak external drive the system is not turbulent; the solution is the linear superposition of non-interacting Gaussian oscillations. However, under significant external drive the solution is the combination of Gaussian fluctuations and nonlinearly interacting coherent structures. (This is true for both the plasma fluctuations and the SST model.)
	\item The system has a mirror symmetry. This means that $K$ -- if expanded in $S$ -- will not have linear terms, as $K$ is invariant under the sign change of the signal while $S$ is not. A more precise formulation of the condition is
	\be
	f\left(x|a\right)=f\left(-x|-a\right),
	\ee
	where $x$ is the measured quantity, $f(x)$ is the probability density function, and $a$ is the dominant parameter \cite{powerlawSK}.
	\item The system also must satisfy several physical requirements (finiteness, smoothness etc.).
\end{enumerate}
Thus the fact that there is a parabolic relation between skewness and kurtosis carries no specific information about the system. The coefficients on the other hand are system specific.


The purpose of this paper is to provide a suitable candidate for the dominant parameter in a wide range of physical systems, despite the fact that the proposed signal-model has been applied only to plasma fluctuations, and to give a more physical meaning to the individual coefficients of the parabolic S-K relation.

The outline of the paper is as follows. In Sec. \ref{sec:model} a general mathematical model is presented to describe signals with intermittent events, for which skewness and kurtosis is calculated. The signal model is applied to plasma physics in Sec. \ref{sec:plasma}, where it is shown that even a simple model for coherent structures -- which are the intermittent events in this case -- can reproduce S-K relations similar to the experimentally observed ones.

\section{Mathematical description of signal}\label{sec:model}

\begin{figure}[H]
\begin {center}
\includegraphics[width=\linewidth]{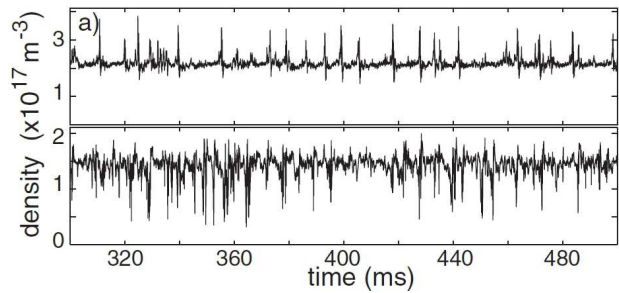}
\caption{Two sample signals from TORPEX \cite{torpex}. Figure reprinted with permission from Ref. 1 (http://link.aps.org/abstract/PRL/v98/e255002), Fig. 1, Copyright 2007 by the American Physical Society.}\label{fig:torpex_signal}
\end {center}
\end{figure}

The measured signals at TORPEX exhibited bursty, intermittent behavior (see Fig. \ref{fig:torpex_signal}). A mathematical description of the signal was proposed by Sandberg et al. in 2009 \cite{Sandberg_PDF}. This univariate model was constructed to be as simple as possible, while exhibiting similar behavior to the experimental signals, thus  in this model the measured signal ($\tilde{Z}$) is
\be
\tilde{Z}=\left(\tilde{X}-\langle\tilde{X} \rangle\right)+\gamma\left(\tilde{X}^2-\langle\tilde{X}^2 \rangle\right),
\ee
where $\tilde{X}$ is a Gaussian random variable. The second term represents the nonlinear part of the signal; its amplitude is set by the $\gamma$ parameter. Although the model did provide a good fit to the experimental signals, it is hard to attribute physical meaning to the nonlinear amplitude $\gamma$, which sets the values of the parabolic coefficients.

	We propose a different mathematical model, where the measured signal ($\tilde{Z}$) is constructed as
\be
\label{eq:signal_model}
	\tilde{Z}=\sum_{i=1}^{\tilde{N}}{\tilde{X}_i}+\tilde{R},
	\ee
	where $\{\tilde{X}_i\}_{i\in [1;\tilde{N}]}$ are independent random variables representing the intermittent events (e.g. turbulent structures), $\tilde{R}$ is the Gaussian background noise with variance $\sigma$ and $\tilde{N}$ is a nonnegative integer valued random variable, which represents the number of intermittent events present simultaneously. Intermittency is achieved if $\tilde{N}$ has a significant probability to take zero value, thus ensuring that intermittent events -- characterized by $\tilde{X}$ -- are detected only during a fraction of the signal.
	
	This model is in accordance with the experimental signals, which show intermittent events superposed on a Gaussian background (see Fig. \ref{fig:torpex_signal}). The experimental signals also show that the variance of the background noise is much smaller than the variance of the events, so we can assume $\sigma^2\ll \lang \tilde{X}^2\rang$, where $\lang ... \rang$ is the expected value operation. 

We also assume that $\tilde{N}$ and $\{\tilde{X}_i\}$ are independent. This means that the number of intermittent events present at the observation point $\left(\tilde{N}\right)$ has no effect on the amplitude distribution of the individual events $\left(\{\tilde{X}_i\}\right)$. Thus we neglect any interaction between them, which is a valid hypothesis as long as the events are rare and the interaction between is not too strong.  This assumption of independence allows us to utilize Wald's equation, which states that $\lang\sum_{i=1}^{\tilde{N}}{\tilde{X}_i}\rang=\lang\tilde{N}\rang\lang\tilde{X}\rang$. As the mean value plays no role in the central moments, we assume it to be zero for brevity $\left(\lang \tilde{Z}\rang=0\right)$, so the central moments of $\tilde{Z}$ are

\begin{eqnarray}
\lang \tilde{Z}^2\rang=\sigma_Z^2=\lang \tilde{N}\rang M_2+\sigma^2 \\
\lang \tilde{Z}^3\rang=\lang \tilde{N}\rang M_3 \\
\lang \tilde{Z}^4\rang=\lang \tilde{N}\rang M_4+3\lang \tilde{N}\left(\tilde{N}-1\right)\rang M_2^2\nonumber \\+3\sigma^4+6\sigma^2 \lang \tilde{N}\rang M_2,
\end{eqnarray}
where $M_i$ is the $i^{th}$ moment of $\tilde{X}$. Although technically only $\sigma^2 \ll M_2$ was specified, this assumption could be extended to $\sigma^2\ll \lang \tilde{N}\rang M_2$, which means that the intermittent events dominate the variance of the signal. The reason $\sigma$ was introduced in the first place, is to explain the experimental results around $[S=0; K=3]$, which we attribute to noise dominated scenarios. 

The definition of skewness ($S$) and kurtosis ($K$) are
\begin{eqnarray}
\label{eq:SK_def}
S=\frac{\lang \left(\tilde{Z}-\lang\tilde{Z}\rang\right)^3\rang}{\sigma_Z^3}\\
K=\frac{\lang \left(\tilde{Z}-\lang\tilde{Z}\rang\right)^4\rang}{\sigma_Z^4},
\end{eqnarray}
where $\sigma_Z$ is the variance of $\tilde{Z}$, which leads to
\begin{eqnarray}
\sigma_Z^2=\lang \tilde{N}\rang M_2 \\
S^2=\frac{M_3^2}{M_2^3\lang \tilde{N}\rang}\\
K=\frac{M_4}{M_2^2\lang \tilde{N}\rang}+3 \frac{\lang \tilde{N}\left(\tilde{N}-1\right)\rang}{\lang \tilde{N}\rang^2},
\end{eqnarray}
resulting in the parabolic relation
\be
\label{eq:parab_fit}
K_{fit}=\frac{M_2 M_4}{M_3^2} S^2 +3 \frac{\lang \tilde{N}\left(\tilde{N}-1\right)\rang}{\lang \tilde{N}\rang^2}.
\ee

While finding the distribution of $\tilde{X}$ is not trivial, as it is likely heavily dependent on the underlying physics (type of system, place of observation), but for the distribution of $\tilde{N}$ an educated guess can be made. These signals experience a discrete number of rare events, which -- in case of plasma turbulence -- can be associated with the emergence of some kind of coherent structures. The simplest model is to assume that the structures are randomly placed, which means that each has a very small chance ($p$) to be in the observed area at a given time. As there are many structures, the number of structures simultaneously present at the observation point can be described as a Poisson process. Therefore it is compelling to take $\tilde{N}$ as a Poisson random variable with expected value $N p$. It is easy to show, that due to the random placement of structures, $N p=f_p$, where $f_p$ is the so called \textit{packing fraction} \cite{chaos_nonlinear}, which is the fraction of the experimental system covered by structures.

In case $\tilde{N}$ follows a Poisson distribution Eq. (\ref{eq:parab_fit}) becomes
\be
\label{eq:parab_fit2}
K_{fit}=\frac{M_2 M_4}{M_3^2} S^2 +3.
\ee
As mentioned before, there is no compelling motivation to use any of the well known statistical distributions for $\tilde{X}$; nevertheless it is likely that it can be sufficiently approximated with one of these, due to the flexible shape of gamma and beta distributions (as demonstrated on TORPEX \cite{torpex}). Using these distributions the parabolic relation of Eq. (\ref{eq:parab_fit2}) takes the form of
\begin{eqnarray}
K_{\gamma}=\frac{3+k}{2+k} S^2 +3, \label{eq:parab_fit_gamma} \\
K_{\beta}=\frac{\left(3+\alpha\right)\left(2+\alpha+\beta\right)}{\left(2+\alpha\right)\left(3+\alpha+\beta\right)} S^2 +3. \label{eq:parab_fit_beta},
\end{eqnarray}
where $k$ is the shape parameter of the Gamma distribution, while $\alpha$, $\beta$ are the two shape parameters of the Beta distribution.

\section{Application to plasma turbulence}\label{sec:plasma}

Plasma turbulence is often investigated using the Hasegawa-Mima model of plasma potential fluctuations. It has been shown \cite{chaos_nonlinear} that the central moments of these potential fluctuations are
\begin{eqnarray}
\label{eq:chaos_sk}
S=\frac{-3 f_p A+f_p A^3}{\left(f_p A^2+1\right)^{3/2}}\\
K=\frac{f_p A^4+6 f_p A^2+3}{\left(f_p A^2+1\right)^{2}}+3,
\end{eqnarray}
where $f_p$ is the packing fraction while $A$ is the amplitude of the vortex solutions (coherent structures, assumed to have Gaussian shape) relative to the linear solutions (background noise). For low $f_p$ the relation between $S$ and $K$ becomes linear, but if we take into account the assumption that the variance is dominated by the vortices ($A^2 f_p\gg 1$), then for small packing fractions we arrive at $K=S^2+3$.

\begin{figure}[H]
\begin {center}
\includegraphics[width=0.9 \linewidth]{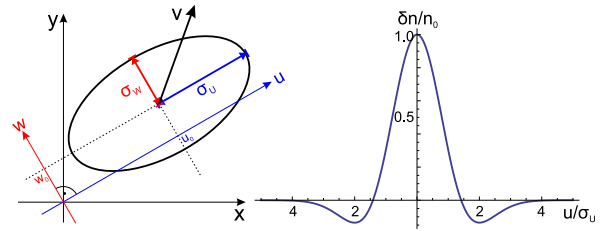}
\caption{(Left) Coordinate system used for the modeling of turbulent structures ($x$,$y$ are the laboratory, while $u$,$w$ are the axes' coordinates), including the velocity ($v$) and scales ($\sigma_U$, $\sigma_W$) of the structure \cite{own_pop}. (Right) Shape of coherent structure along its axis. }\label{fig:model}
\end {center}
\end{figure}

Meanwhile, for our analysis of turbulent measurements we develop a different Gaussian model for the coherent structures \cite{own_pop}, where -- similarly to the previous analysis -- we assume that the fluctuation of the plasma density is composed of small coherent structures. The main difference between this model and the one used in the derivation of Eq. \ref{eq:chaos_sk} is the shape of the coherent structures. Horton and Ichikawa assumed a simple Gaussian shape \cite{chaos_nonlinear}, while we argue that since turbulent dynamics conserve particles, the net change in particle number due to the presence of a structure must be zero. Thus we assume that these structures have polynomial times Gaussian spatial distributions (see Eq. (\ref{eq:better_model})) in the direction of both their axes (see Fig.\ref{fig:model}) and a Gaussian time decay. The model also assumes that these move at a constant velocity and have the same size and orientation. We thus have
\begin{eqnarray}
\label{eq:better_model}
\delta n_i(u,w,t)=e^{-\hat{u}_i^2-\hat{w}_i^2-\hat{t}_i^2}(1-\hat{u}_i^2-\hat{w}_i^2) \\
\delta n=\sum_{i=1}^{N}{\delta n_i},
\end{eqnarray}
where $\delta n_i$ is the density perturbation caused by the $\rm i^{th}$ coherent structure, $\delta n$ is the total density perturbation and $N$ is the number of structures, while $\hat{u}^2$ is the normalized version of $u^2$, defined as $\hat{u}_i^2=\left(u-u_0(i,t)\right)^2/\left(2\sigma_u^2\right)$, where $u_0(i,t)=u_{i}+v_u t$ thus $\hat{u}^2$ is essentially a Gaussian exponent with a moving center (for further details see Fig. \ref{fig:model}).

It should be noted that this model not only belongs to the signal group described by Eq. (\ref{eq:signal_model}), but for a large number of independent structures ($N\gg 1$) the temporal distribution of events $\left(\tilde{N}\right)$ follows a Poisson distribution -- as mentioned in Sec. \ref{sec:model} -- which leads to a parabolic relation of the form of Eq. \ref{eq:parab_fit2}.

As the signal $\tilde{Z}$ is proportional to the local density perturbation $\delta n$, it is enough to calculate the moments of the latter, as their central moments are the same. Using Eq. (\ref{eq:SK_def}) these central moments can be calculated (see Appendix \ref{app: mom_calc} for details). Assuming that a large number of structures are present ($N\gg 1$) the results take the simple form of
\begin{eqnarray}
\langle \tilde{Z} \rangle =\langle \delta n \rangle= 0 \label{eq:model_expval}\\
\sigma_{\delta n}= \frac{\pi^{1/4}}{4}\sqrt{f} \\
S=\frac{64 \sqrt{2}}{27\sqrt{3} \pi^{1/4}}\frac{1}{\sqrt{f}} \\
K=3+\frac{15}{2 \sqrt{2\pi}} \frac{1}{f}\label{eq:model_K},
\end{eqnarray}
where $f=8\pi\sigma_T\sigma_U\sigma_W/(\Delta T\Delta U\Delta W)$ is the \textit{filling value} -- which is the expected number of structures present at an arbitrary point in space and time -- while $\Delta T, \Delta U, \Delta W$ are the time length of the measurement and the size of the poloidal area in which the structures are randomly distributed. This means $f$ is equivalent to $\langle \tilde{N}\rangle$, when compared to the generic model of Eq. (\ref{eq:signal_model}). The filling value is also an equivalent of the \textit{packing fraction} ($f_p$), but generalized to account for the temporal decay of perturbations. Experiments have shown that this value is low ($\ord\left(10\%\right)$ \cite{fillingexp}), making the contribution from coherent structures the dominant term in the skewness and kurtosis. This leads to the following parabolic relation:
\be
\label{model_anal_SK}
K\approx 1.41 S^2+3,
\ee
which is very similar to the experimental relation of Eq. (\ref{eq:exp_parabola}). 

To validate the previous calculation a succession of simulations were run, where randomly distributed coherent structures -- which have the shape and time evolution described in Eq. (\ref{eq:better_model}) -- propagate in a 2D plane. During these simulation runs the background noise was kept constant and a series of signals were simulated for different filling values. As the sign of the skewness is determined by the sign of the density perturbations, each simulation was carried out for either positive of negative density perturbations to make visual comparison with the experimental S-K relation (Fig. \ref{fig:torpex_SK}) easier. This is further motivated by the fact that most experimental signals were dominated by either positive or negative perturbations (see Fig. \ref{fig:torpex_signal}) \cite{torpex}. At this point we would like to note that the presence of both negative and positive perturbations could account for the experimental data points with significant kurtosis around S=0.

\begin{figure*}
\begin {center}
\includegraphics[width=\textwidth]{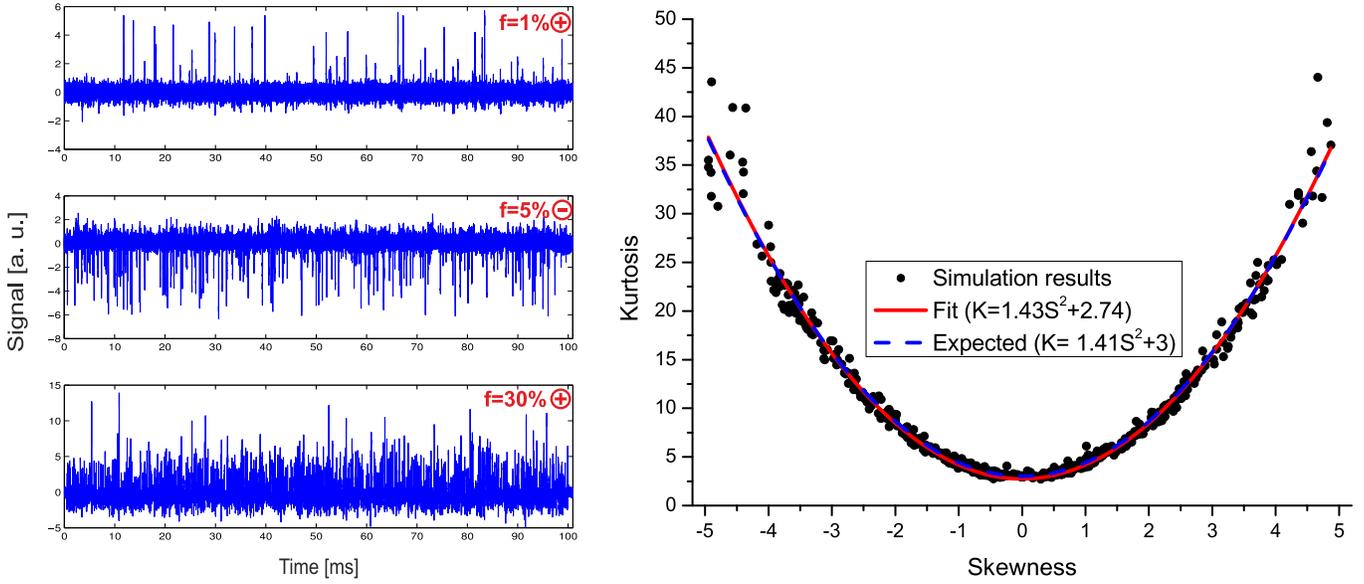}
\caption{Sample simulated signals (left) and skewness and kurtosis values (right) for various filling values with constant noise. The left figures show the signal forms for different filling values and perturbation signs ($+$ or $-$). In the right figure the solid blue line shows the fitted parabola, while the solid green line is the expected parabola according to Eq. (\ref{model_anal_SK}).}\label{fig:SK_signal_fig}
\end {center}
\end{figure*}

Fig. \ref{fig:SK_signal_fig} shows the skewness and kurtosis values obtained from the simulation along with some sample signals, and it is apparent that this result somewhat resembles the experimental results from TORPEX (Fig. \ref{fig:torpex_SK} and \ref{fig:torpex_signal}). It should be noted that points close to $[S=0; K=3]$ are the result of very low filling values, where the signal is dominated by noise. Although very high filling values  ($f\gg 1$) could also reproduce this effect but these scenarios are outside the scope of this model, as they would involve the overlap of intermittent events, in which case the interaction between them could not be neglected. It is also very important to note that most of the TORPEX measurements were done in the SOL, where coherent structures can be highly asymmetric and have non-Gaussian shapes, so we do not expect to find a perfect fit for Eq. (\ref{eq:exp_parabola}) with a Gaussian model.

\section{Conclusion}

In summary, Sattin et al. stated that the parabolic relation in itself contains no deep physical meaning but it is the manifestation of an intrinsic mirror symmetry \cite{Sattin}.


The present work has shown that the parabolic relationship is a manifestation of the dominant influence of a single parameter, namely the fill factor ($f$) of the fluctuation statistics observed at various locations in a fusion plasma and possibly other systems as well. The physical interpretation is the following. It is known that plasma turbulence signals are close to Gaussian statistics inside the Last Closed Flux Surface (LCFS) \cite{gauss_zoletnik}, while they exhibit increasingly non-Gaussian statistics as one moves outward through the Scrape-Off Layer (SOL). This can be understood as changing intermittency, that decreases $f$ towards the SOL. If the statistical moments of the individual events are not varying drastically across different locations then $f$ changes $S$ and $K$ in such a way that the parabolic relationship is kept as shown by Eq. (\ref{eq:parab_fit}). The decrease of $f$ can be understood, if one assumes that turbulence events originate around the LCFS and move outwards as seen in the experiments. Larger structures live longer, therefore smaller number of events reach the outer SOL, thus the decreasing the fill factor.

In addition, our model allows us to attribute a meaning to the coefficients of the parabolic relation $\left(K=aS^2+b\right)$, as the offset $b$ is only dependent on the temporal statistics of the events $\tilde{N}$, while the quadratic term $a$ depends only on the spatial distribution of individual events $\tilde{X}$, as shown in Eq. \ref{eq:parab_fit}. For independent events the temporal statistics can be approximated as a Poisson process, setting the offset at 3, so if the experimental value of 2.78 in plasmas can be considered significantly different from 3, then it might indicate a deviation from Poisson statistics, which implies a temporal correlation of events (e.g. interaction between coherent structures). Meanwhile, the structure term ($M_2 M_4/M_3^2$) is not necessarily constant in time, which could explain the sometimes significant deviations from the fitted parabolic relation in the experiment (Fig. \ref{fig:torpex_SK}). Another possible explanation is the simultaneous presence of positive and negative perturbations, which increase kurtosis without significantly affecting skewness.


The observation of the same S-K relationship in sea temperature fluctuations \cite{SS} might be a manifestation of similar statistical properties, where intermittent weather events affect the sea surface temperature with varying frequency. Attempting to create a physically motivated model for this problem is outside the scope of this paper. However, it should be noted that intermittency already plays an important role in statistical fluid models describing the atmosphere, where it is often used to describe advection \cite{chatwin_advances,intermitt_factor}. In a sense these models can be taken as a subtype of the model proposed by this paper (Eq. (\ref{eq:signal_model})), as the introduced $\gamma$ intermittency factor has the same effect as setting $\tilde{N}$ as a Bernoulli distribution with $\gamma$ as the main parameter. These models also lead to parabolic S-K relations \cite{chatwin_advances}.

\appendix

\section{Calculation of higher moments for Gaussian coherent structure model}\label{app: mom_calc}

For the central moments of the Gaussian coherent structure model in Sec. \ref{sec:plasma} the higher moments of the density fluctuation must be calculated. In this section the derivation leading to Eqs. (\ref{eq:model_expval})-(\ref{eq:model_K}) is detailed.

According to Eq. (\ref{eq:better_model}) the local fluctuating density ($\delta n$) is defined as the sum of the local contributions from each coherent structure, so the $\rm{k^{th}}$ moment of the density is
\be
\label{eq:kth_mom}
\lang \delta n^k \rang =\sum_{\{i_1, i_2 ... i_k\}}{\lang \delta n_{i_1}\delta n_{i_2}...\delta n_{i_k}\rang}.
\ee
For brevity let us introduce the following quantities:
\begin{eqnarray}
M_k(N)\equiv\lang \delta n^k \rang=\lang \left(\sum_{i=1}^{N}{n_i}\right)^k\rang, \\
E_k\equiv\lang \delta n_i^k \rang=\lang \left(e^{-\hat{u}_i^2-\hat{w}_i^2-\hat{t}_i^2}(1-\hat{u}_i^2-\hat{w}_i^2)\right)^k \rang,
\end{eqnarray}
where $M_k(N)$ is essentially the $\rm{k^{th}}$ moment of the density defined in Eq. \ref{eq:kth_mom} if $N$ structures are present, while $E_k$ is the $\rm{k^{th}}$ moment of the density perturbation created by a single coherent structure. Also
\begin{widetext}
\be
\lang \delta n_i^k \rang=\frac{1}{\Delta U \Delta W \Delta T}\int_{-\Delta U/2}^{\Delta U/2}{\int_{-\Delta W/2}^{\Delta W/2}{\int_{-\Delta T/2}^{\Delta T/2}{ \delta n_i^k du_i}\rm dw_i} \rm dt_i}\approx 
\frac{1}{\Delta U \Delta W \Delta T}\int_{-\infty}^{\infty}{\int_{-\infty}^{\infty}{\int_{-\infty}^{\infty}{ \delta n_i^k du_i}\rm dw_i} \rm dt_i},
\ee
\end{widetext}
where we expand the integral limits to infinity, which is a valid approximation as long as the lifetime $\sigma_T$ and spatial extents $\sigma_U$, $\sigma_W$ of the coherent structures are much smaller than the time length $\Delta T$ of the measurement and the spatial extents, $\Delta U$, $\Delta W$ of the poloidal area. Now $E_k$ can be easily calculated, as for any value of $k$ $\delta n_i^k$ takes the form of $\rm{Gaussian}\times \rm{polynomial}$ and thus $\lang \delta n_i^k \rang$ is a sum of Gaussian integrals, which can be easily evaluated.

Let us try to find a general formula for $M_k(N)$. The first moments are
\begin{eqnarray}
\label{eq:first_moms}
M_1=N E_1, \nonumber\\
M_2=N E_2+N(N-1)E_1 \nonumber \\
M_3=N E_3+3 N(N-1)E_2 E_1+N(N-1)(N-2) E_1^3 \nonumber \\
M_4=N E_4+4 N(N-1)E_3 E_1 + 3 (N (N-1) E_2^2+ \nonumber \\ 6 N(N-1)(N-2) E_2 E_1^2+ N(N-1)(N-2)(N-3) E_1^4 \nonumber \\
... \nonumber \\
\end{eqnarray}
With the use of some combinatorics a recursion formula can be found for $M_k(N)$:
\be
\label{eq:recursion}
M_k(N)=\sum_{j=1}^{k}{\binom{k-1}{j-1} N E_j M_{k-j}(N-1)}.
\ee

We also have
\be
E_1\propto\int{\int{ \int{e^{-\hat{u}^2-\hat{w}^2-\hat{t}^2}(1-\hat{u}^2-\hat{w}^2) dt_i} dw_i} du_i}=0,
\ee
as required by particle number conservation, which leads to
\be
\lang \delta n \rang=M_1=N E_1=0.
\ee
$\sigma_{\delta n}$ can be calculated using $E_2$, which is
\begin{widetext}
\be
E_2=\frac{1}{\Delta U\Delta W\Delta T}\int{\int{ \int{e^{-2\hat{u}^2-2\hat{w}^2-2\hat{t}^2}(1-\hat{u}^2-\hat{w}^2)^2 dt_i} dw_i} du_i}=\frac{\pi^{3/2}}{2}\frac{\sigma_U \sigma_W \sigma_T}{\Delta U\Delta W\Delta T},
\ee
\end{widetext}
thus the variance is
\be
\sigma_{\delta n}^2=M_2(N)-M_1(N)^2=N E_2=\frac{\pi^{1/2}}{16} f,
\ee
where we used the definition of $f$ from Sec. \ref{sec:plasma}. Similarly $E_3$ and $E_4$ can be calculated:
\begin{eqnarray}
E_3=\frac{8\sqrt{2}}{27 \sqrt{3}}\frac{\pi^{3/2}\sigma_U \sigma_W \sigma_T}{\Delta U\Delta W\Delta T}\\
E_4=\frac{15}{64 \sqrt{2}}\frac{\pi^{3/2}\sigma_U \sigma_W \sigma_T}{\Delta U\Delta W\Delta T}.
\end{eqnarray}
Taking the $N\rightarrow\infty$ limit simplifies $S$ and $K$ as all factors of form $\left(N-k\right)$ become $N$, allowing us to simplify the expressions as
\begin{eqnarray}
S=\frac{64 \sqrt{2}}{27\sqrt{3} \pi^{1/4}}\frac{1}{\sqrt{f}} \\
K=3+\frac{15}{2 \sqrt{2\pi}} \frac{1}{f}.
\end{eqnarray}


\end{document}